\documentclass[graybox]{svmult}
\usepackage{mathptmx}       
\usepackage{makeidx}         
\usepackage{graphicx}        
\begin{document}

\title{Shape models and physical properties of asteroids}
\author{Santana-Ros T., Dudzi{\'n}ski G. and Bartczak P.}
\institute{Santana-Ros T. \at Astronomical Observatory Institute, Faculty of Physics, Adam Mickiewicz University, S{\l}oneczna 36, 60-286 Pozna{\'n}, Poland, \email{tonsan@amu.edu.pl}
\and Dudzi{\'n}ski G. \at Astronomical Observatory Institute, Faculty of Physics, Adam Mickiewicz University, S{\l}oneczna 36, 60-286 Pozna{\'n}, Poland \email{g.dudzinski@amu.edu.pl}
\and Bartczak P. \at Astronomical Observatory Institute, Faculty of Physics, Adam Mickiewicz University, S{\l}oneczna 36, 60-286 Pozna{\'n}, Poland \email{przebar@amu.edu.pl}}
%
%
\maketitle

\abstract{Despite the large amount of high quality data generated in
recent space encounters with asteroids, the majority of our knowledge
about these objects comes from ground based observations.
Asteroids travelling in orbits that are potentially hazardous for the Earth form
an especially interesting group to be studied. In order to predict their
orbital evolution, it is necessary to investigate their physical
properties. This paper briefly describes the data requirements and
different techniques used to solve the lightcurve inversion problem.
Although photometry is the most abundant type of observational data,
models of asteroids can be obtained using various data types and
techniques. We describe the potential of radar imaging and stellar
occultation timings to be combined with disk-integrated photometry in order to
reveal information about physical properties of asteroids.}

\section{Introduction}
\label{sec:1}

Asteroids play an important role in the formation and evolution models
of the Solar System and have a direct link to life on Earth. They are
connected to the delivery of water and probably also organic material to
our planet and therefore are crucial for the development of life. On the
other hand, some of them are considered as potentially hazardous for our
future.

News about the discovery of a new hazardous asteroid appear regularly in the
media. The last popular case was the close fly-by on 30th October 2015
of asteroid 2015 TB145 (the so-called "Halloween asteroid"). However, the news coverage usually tends to be 
sensationalist rather than scientific, as these asteroids don't represent an imminent risk of impact on Earth. The identification of a
potential hazard comes from the forward integration of the asteroid's motion and evolution of its orbit, 
and the calculation of a probability of an impact (usually less than one in a few
thousand chance) with our planet over the next decades. An example of
these predictions can be found in the Sentry Risk Table
(\url{http://neo.jpl.nasa.gov/risks/}) maintained by NASA's Jet
Propulsion Laboratory (JPL). In order to obtain the best possible
accuracy in these calculations, there are two crucial actions to be
taken: (1) a very precise determination of the orbital parameters by
astrometric measurements and (2) a study of the physical properties of
the body. 

The reason for the first action is obvious. Asteroids approaching the
Earth have their orbits modified due to gravitational interactions and therefore
regular astrometric measurements are required to constrain the orbital
parameters. The better the orbit of an asteroid is defined, the greater
will be the accuracy of our future predictions. However, for longer term
predictions (i.e. several decades), second-order effects such as
nongravitational forces plays a key role on the evolution of the orbit. The
most important nongravitational perturbation is caused by the Yarkovsky effect
(\cite{opik}, \cite{neiman}) which is due to radiative recoil of anisotropic thermal
emission and causes asteroids to undergo a secular semimajor axis drift.
The Yarkovsky acceleration depends on several physical quantities such
as spin state, size, mass, shape and thermal properties \cite{vokro}. On
the other hand, the Yarkovsky--O'Keefe--Radzievskii--Paddack (YORP)
effect slowly modifies the spin rate of asteroids with irregular shapes,
which in turn affects the Yarkovsky acceleration rate. Rubincam
\cite{rubi} showed that YORP is strongly dependent on an asteroid's
shape, size, distance from the Sun, and orientation. 

Unlike asteroid's distance from the Sun, which can be trivially
calculated with good accuracy when knowing the orbital parameters,
deriving other physical properties of the asteroid, like spin state or
shape, requires advanced inversion techniques and well-planned
observations with a good coverage of various viewing geometries. Relative photometry is, by
far, the richest source for deriving asteroid models. However, other
observing techniques -- such as Doppler-delay radar imaging, adaptive
optics or stellar occultations -- can provide valuable information of
the asteroid's shape and, more importantly, they allow for the model to
be scaled.  

In this paper we briefly review the importance of investigating the
shapes and spin states of asteroids in the context of identifying
potentially hazardous bodies. We describe the requirements of
photometric data in order to solve the inversion problem, and we discuss
different shape solutions. In the last chapter, we review the potential
of combining lightcurves with other data types.

\section{The importance of asteroid modelling in the assessment of asteroid impact hazard}
\label{sec:2}

The evolutionary processes that asteroids undergo have been traditionally explained
by gravitational perturbations (e.g. gravitational pulls in close encounters with other bodies, orbital
resonances) and collisions between these small bodies. The classical
asteroid evolution model has been useful to explain how asteroid
populations have evolved with time. Particularly interesting was efforts to
understand the main source of Mars-crossing and Near Earth Object (NEO)
populations, which we believe are fed with material from the main
belt delivered by the effects of secular resonances (\cite{wet}, \cite{wis}). 
However, classical models are unable to explain some of the
physical characteristics observed in the NEO population. For instance,
according to the classical model, the only processes able to inject these
bodies into orbital resonances are asteroid collisions. Meteoroids
delivered through this process should present cosmic-ray exposures (CRE) - a measure of their ages -
of the order of million years \cite{glad}. However, the observed CRE for
NEO population are hundreds of times higher. 

In turn, these high CRE values can be well explained when introducing
nongravitational effects to the evolution model, because such effects can result in A slow delivery of 
material to orbital resonance zones. Specifically,
the Yarkovsky effect induces a tiny force to small bodies by the
reradiation of sunlight in the form of thermal energy. This force slowly
changes the object's semimajor axis, changing it's orbit inwards (for
objects rotating with retrograde sense) or outwards (prograde sense of
rotation) with respect to the Sun. Yarkovsky effect is divided into two types of perturbations:
(1) a diurnal perturbation due to the body rotation and (2) a seasonal
perturbation that depends on the heliocentric longitude of the object.
The acceleration $da/dt$ for each perturbation is given by the following
equations (see \cite{bottke} for further details):\\

\begin{equation}
\bigg( \frac{da}{dt} \bigg)_{diurnal} = -\frac{8\alpha}{9}\frac{\Phi}{n}F_{\omega}(R,l_{v},\Theta)\cos\gamma +	\mathcal{O} (e)
\end{equation}\\

\begin{equation}
\bigg( \frac{da}{dt} \bigg)_{seasonal} = \frac{4\alpha}{9}\frac{\Phi}{n}F_{n}(R,l_{v},\Theta)\sin^2\gamma + \mathcal{O} (e)
\end{equation}\\

where $\alpha$ is the albedo-factor, $\Phi$ is the radiation pressure
coeficient and $\gamma$ is obliquity of the spin axis. The function $F(R,l_{v},\Theta)$ depends on the radius of the body $R$, the penetration depth $l_{v}$ and the
 thermal parameter $\Theta$ (see the explicit form of this function in \cite{vokro}). The total
acceleration is the superposition of the diurnal and seasonal terms. Thus, the
magnitude of the Yarkovsky effect depends on the object's distance from
the Sun, the spin axis orientation, and the body's physical
characteristics (i.e., size, shape, thermal properties, and rotation
period).

On the other hand, another nongravitational effect called YORP, is
capable of modifying the spin rates and axis orientations of asteroids.
Reemitted photons apply a recoil force $d\mathbf{f}$ normal to the
surface. If the body is not perfectly symmetric, the sum of these forces
produces a thermal torque (see \cite{bottke} for further details):\\

\begin{equation}
\mathbf{T} = \int{\mathbf{r} \times d\mathbf{f}}
\end{equation}\\

where $\mathbf{r}$ is the position vector of a surface element $d\mathbf{S}$.

In this case, the effect strongly depends on the body's shape (i.e. it's
irregularities), and to calculate the effect it is necessary to model
the body's surface temperature distribution (see for instance \cite{dobro} or
\cite{vokrocapek}). 

Thus, in order to include these nongravitational effects in the orbital
calculations it is necessary to know in detail the physical properties of
the asteroid. For NEOs the most commonly used technique to obtain THE body's
size is radar ranging (see Section 4 for further details). Surface 
thermal properties are related to the roughness of the body surface and its
regolith depth. Such properties can be derived, for instance, using infrared
interferometric observations. For modelling the Yarkovsky effect, it is
essential to know the asteroid's spin state and its axis orientation.
A convex representation of the body shape is usually
enough to solve the lightcurve inversion problem, what is the main
source of asteroid models. However, as the YORP effect is very sensitive
to irregularities of the body shape, a high resolution shape model is
required to calculate this effect. In this sense, shape models obtained
by spacecraft in situ measurements represents the ideal case. Obviously,
this kind of observations are limited to a bunch of asteroids which have
been visited by spacecrafts, therefore generally we have to rely on remote
observations. Radar echo can be useful to retrieve a complex shape
model, including concavities. However, shapes obtained with this
technique are not always reliable, and care should be taken when
deriving results from this technique alone. Moreover, before deriving
the shape from the radar echo, it is necessary to know the asteroid's
spin axis orientation. Lightcurves are a great source of information regarding the
asteroid's rotational state. As the Sun-asteroid-observer geometry
changes so does the observed lightcurve. If the observations are
gathered at a variety of geometries (see Section \ref{sec:lightcurves}
for requirements) it is possible to reconstruct a shape and spin of an asteroid.
In the next chapter we describe the data required to solve the inversion
problem, as well as the shape representations commonly in use.

\section{Models based on photometry}
\label{sec:3}

Deriving asteroid's spin state and shape is a necessity in order to model the nongravitational forces. To
that end, photometry is by far the most fruitful observing
technique. The classical photometric observations of asteroids
(henceforth "dense lightcurves") that have been collected during the
last decades, are the main source of our knowledge about asteroids and
their physical parameters. However, gathering enough photometric data to
derive a model is an arduous task, which requires good planning and,
often, a collaboration between several observers. When the collected
data fulfils the requirements, an inversion technique can be applied to
obtain a model of an asteroid. Such model includes asteroid's
rotational state as well as an approximation of the shape of the body.
In this sense, different shape representations can be used (e.g. triaxial
ellipsoid, convex or nonconvex figures). In this chapter we summarize the
modelling process, from the obtention of data, to finding the solution of the
inversion problem.

\subsection{Requirements for the lightcurves} %
\label{sec:lightcurves}

Lightcurves can be obtained by comparing the apparent brightness of an
asteroid with that of comparison stars (relative photometry), or with
that of photometric standard stars (absolute photometry). It might seem
that performing absolute photometric measurements should always be
preferred. However, absolute photometry limits observable targets 
to brigh asteroids, due to the use of filters, not to mention the
requirement of excellent weather conditions. These constraints are of
special importance when observing asteroids with small amplitude range
(e.g. below 0.1 mag), as uncertainties of absolute magnitude measurements can
be of a comparable range.

\begin{figure}
\centering
\includegraphics[width=110mm]{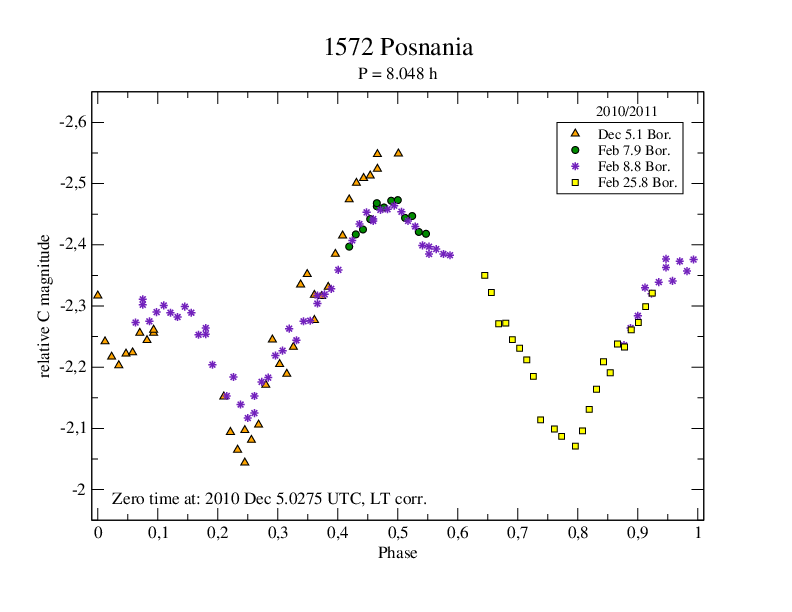}
\caption{Composite lightcurve of asteroid (1572) Posnania observed with a 0.4m Newtonian telescope at the Borowiec observatory.}
\label{lc}
\end{figure}

The usual lightcurve format for one apparition (the period during which
the asteroid is observable from the Earth) is basically a series of 
photometric measurements collected during several observing nights
(e.g. Fig.~\ref{lc}), with a 0.05 magnitude precision at worst. Ideally,
a lightcurve should contain at least 50 well placed data points with a
precision better than 0.01 magnitude. A general practice is to do continuous asteroid exposures 1 to 5
minutes long, depending on the object brightness. The field of
view (FOV) should be large enough to include three comparison stars of brightness similar to that of an asteroid
and, preferably, also of similar colour. When the lightcurve is
complete, the rotational period of the asteroid is well covered but the
quantity of information on the body shape is limited, as the viewing geometry
of the asteroid is almost constant during observations. 
Consequently, to obtain a unique spin and shape solution, we need
a set of dense lightcurves observed at a large span of viewing
geometries (i.e. well-spread ecliptic longitudes and a substantial span of phase
angles). This observational constraint makes this technique highly time
consuming, what is significantly limiting the number of objects for
which we have enough dense lightcurves to derive a complex shape of the
body.

The quality of photometric observations is related to signal-to-noise ratio
(SNR), which is a statistical term that defines the ratio between the
useful signal (photons from the object) versus the total signal received
(photons from the object, sky background, inherent noise in the chip,
etc). The larger this number, the more signal (photons) from the
target or comparison star only. An ideal SNR value would be above 100, which means
that the noise is about 1 per cent of the total signal, or in other
words, that the measurements are of about 0.01 mag precision. In
practice one can still get good results when the SNR drops to 50 or 30.
Getting the necessary SNR depends on many factors: size of the telescope, type
of CCD camera, whether or not filters are used, the sky background brightness, how
fast the asteroid is moving, the quality of dark and flat frames,
etc.

\subsection{Shape models}
\label{sec:shape}

Long term variations in apparent brightness of an asteroid depend
mainly on its distant to the Sun and to the observer, and the angle
between their pointing vectors (the so-called phase angle). However, any
asteroid with non-spherical shape (practically each asteroid) has also shorter
cyclical variations due to its rotation. The lightcurve characteristics
(e.g. amplitude, period, shape) depends on the asteroid's spin state,
but also on its shape. If observed in equatorial view, an elongated body
will produce a lightcurve with large amplitude, while a nearly spherical
object will present a lightcurve with low amplitude. However, if the
observation is taken in a pole-on viewing geometry, its lightcurve will
be almost flat, no matter the body's shape.

In order to reproduce such variations, the inversion method has to
include a recreation of the real asteroid's shape. A 3-axis ellipsoid
shape can be a fairly good approximation to solve the inversion problem
for the majority of cases (\cite{conos}, \cite{tadeusz}, \cite{torppa}).
Such ellipsoid can be defined as the region bound by a surface given by
the equation:

\begin{eqnarray}
(x/a)^2+(y/b)^2+(z/c)^2=1
\end{eqnarray}

where $a$, $b$ and $c$ are the semi-axes and satisfy the condition $a
\geq b \geq c$. 

Most of the asteroids observed at low phase angles show two maxima and two minima per rotational
cycle. Such a lightcurve can be explained considering a ellipsoidal
shape rotating with a given sidereal period ($P$) about its spin axis 
(which orientation is described by its north pole position the sky in the ecliptic coordinates $\lambda$, $\beta$) 
which usually is an axis of the biggest moment of inertia. The shape of the
ellipsoid is then defined by two parameters, namely, the ratios of the
lengths of the principal axes ($\frac{a}{b}$ and $\frac{b}{c}$). A model
based on such a representation of shape is completed with an initial
rotation angle $\phi_0$ and the sense of rotation of the body (prograde
or retrograde, defined by a sign of $\beta$). Using these parameters, it is possible to explain the
variation in brightness of an asteroid, not only due to a rotation
itself, but also due to the changes of the viewing geometry for the
Sun--observer--asteroid system. Analytically, the brightness of the
asteroid at a given time $t$, is proportional to the surface area seen
from a given reference frame (cross-section of the asteroid presented to
the observer). The cross-section can be calculated using the following
equation:

\begin{eqnarray}
S=\pi abc \frac{\cos^2{\phi}\sin^2{\gamma}}{a^2}+\frac{\sin^2{\phi}\sin^2{\gamma}}{b^2}+\frac{\cos^2{\gamma}}{c^2}
\end{eqnarray}

where $\phi$ is the asteroid's rotation angle and $\gamma$ is the aspect
angle (the angle between the rotation axis and the asteroid--observer
line of sight). As we change the rotation angle $\phi$, so does the
cross-section observed, thus we obtain a sinusoidal variation of 
brightness.

Nevertheless, some lightcurve shapes cannot be explained by the use of
a simple triaxial ellipsoid model. Asteroids with complex shapes can
produce lightcurves with 3 or more maxima per cycle. In the majority of cases
these asteroids are modelled using a convex representation of their real
shapes (\cite{kaasa01}, \cite{kaasaetal}), which despite being a
first-order approximation of the real shape of the body, have been
proven to be good enough to fit the lightcurves and to derive asteroid's
main physical parameters. In short, this method attempts to fit a set of
parameters namely:

\begin{itemize}
\item A convex shape represented as a collection of triangular facets
\item Sidereal rotation period
\item Pole direction
\item Albedo-dependent coefficient for Lommel-Seeliger and Lambert scattering laws
\end{itemize}

The standard solution of the inversion problem involves minimizing
the residuals between disk-integrated photometric data and synthetic
brightness generated by the model. The process relies on the Minkovski
minimization stability of convex bodies (\cite{lamb}) which makes the
method not very sensitive to random noise in data. This inversion
technique has been used by several authors during the last decade (e.g.
\cite{durech}, \cite{ania11}, \cite{ania12}) resulting in around a hundred
of convex asteroid models based on dense lightcurves.

However, from direct images of asteroids obtained by radar, adaptive
optics or during space missions like NEAR Shoemaker or Hayabusa, we know
that the real shapes of asteroids are not convex, but generally are full
of concavities. In order to obtain a more accurate (realistic) shape
model, alternative methods have been proposed. For instance, Bartczak
et al. \cite{BartczakAntiope} recently developed a new inversion method
called SAGE (Shaping Asteroids with Genetic Evolution) capable to derive
nonconvex shape models for single and binary asteroids relying on their
disk-integrated photometric measurements. In this case, the optimization
problem is tackled by a genetic algorithm, which randomly mutates the
model parameters and selects the best trial solutions until the
evolution stabilizes. These models confirm the pole directions and
rotation periods derived with previous methods, and additionally 
highly detailed description of the asteroids' shape allows more
accurate determinations of their physical properties, like the volume and
in turn, density.

In all the cases, the inversion techniques generally relies on relative
photometry, so the resulting models are also relative in terms of
dimensions. In order to scale them, we need an absolute measurement of
the asteroid size. This can be obtained from other observation
techniques like the time chords recorded during a stellar occultation by
an asteroid, or direct imaging techniques, like radar (see Section
\ref{sec:4}).

Several approaches to the multi-data inversion problem have been developed
during the last years. For instance, the KOALA (Knitted Occultation,
Adaptive optics and Lightcurve Analysis \cite{carry}) algorithm
solutions are based on lightcurves and AO silhouette contours, while ADAM
(All-Data Asteroid Modeling \cite{adam}) is a collection of functions
from which one can tailor an inversion procedure for multiple data
sources including direct imaging, radar and interferometry.

For all the methods described, the main constraint for enlarging the
number of derived models is the availability of good-quality photometric
data fulfilling the requirements described in Section \ref{sec:2}. The
organization of observing campaigns, can potentially generate enough
dense photometric data to derive a few tens of new models per year.

On the other hand, during the last years, some observatories around the
world have conducted sky surveys mainly focused on detecting new NEAs or to
improve their orbital parameter (e.g. USNO in Flagstaff, or Catalina
Sky Survey). As a by-product of these astrometric survey
programs a vast amount of sparse-in-time photometric measurements for
tens of thousands of asteroids have been retrieved. For each object some
tens, or often hundreds of discrete observations were collected for different 
geometries and illuminations. Combining these datasets with dense
lightcurves allowed some authors to increase the modelled population of
asteroids from 100 (classical photometry) to 400 (combination of
classical and sparse photometric data), using a modified version of the
convex lightcurve inversion method (e.g. \cite{hanus11},
\cite{hanus13}). The resulting models resemble the ones obtained with
dense lightcurves (are equivalent in terms of spin solution) but the
shape model is usually a low-resolution, "angular" convex shape, due to the limited
quality of the data. 

In turn, \textit{Gaia} observations will generate a similar sparse set
of photometric measurements during its 5 years operation. But the data
improvement will be considerable, both in terms of quantity
(observations are expected for $\sim$300.000 asteroids, \cite{mignard})
and quality (the photometric accuracy is estimated to be $\sim$0.01 mag
for asteroids up to 18 magnitude, and $\sim$0.03 mag up to 20 magnitude
\cite{cellino06}, \cite{cellino09}). As a result of this enormous amount
of new data, asteroid models for at least 10.000 objects are expected. This means an improvement of two order of magnitudes from our current knowledge level. 

However, on average \textit{Gaia} will observe each asteroid ~50-70
times during 5 years. Despite of the high data quality, this number of
measurements is not enough to constrain a complex model shape by its
own. Moreover, processing such enormous amount of data would be highly
CPU demanding. For these reasons, the inversion method chosen for
inverting \textit{Gaia} photometric data of asteroids is a low CPU
demanding method: a triaxial ellipsoid model, which brightness can be analytically
calculated as a function of the asteroid--Gaia--Sun position vectors and
the Lommel-Seeliger law \cite{cellino15}. This method, while simple, has
been proved to be effective even when inverting synthetic data generated
with nonconvex shapes \cite{toni}. The results coming out from
\textit{Gaia} are expected to have a direct impact on the Solar System
formation theories, as a statistically large sample of objects with
known properties may reveal physical effects which play an important
role for the whole population.

It is worth noting that even such precise data will provide models not completely free from various biases, 
or selection effects, that favour i.e. elongated targets with extreme obliquities of spin axes \cite{toni}. 
It is important for ground-based studies to focus on those targets that will not be fully covered by studies  
based on data from \textit{Gaia} or other future surveys.

\subsection{Models of binary asteroids}
\label{sec:binary}

One particularly interesting case are the asteroids with satellites.
Such systems are specially appreciated by the Solar System researchers
as they give a unique opportunity to derive the mass of the components
directly from the third Kepler's law. For this reason, they are
invaluable targets for studies on internal structure and composition.

The synchronous binary systems have been extensively studied and modelled (e.g. in \cite{tadeusz02}, 
\cite{tadeusz04} and \cite{agnieszka}). Recently a new
algorithm capable to generate model solutions for binary asteroids has
been developed using a nonconvex shape representation of the components
\cite{BartczakAntiope}. As the model is able to reproduce body
concavities, the relative volume obtained for the components is more
accurate than for the previous models, which were based in Roche
ellipsoids \cite{descamps}, having a direct impact on the density
calculation.

We currently know of more than a hundred binary asteroids in the main
belt, and about three hundred in total if binary NEOs and TNOs are included. The
majority of them have been discovered by recording their mutual events
in a classical dense lightcurve. Resolved imaging such as the ones
obtained from radar or adaptive optics have allowed to confirm or, in a
few cases, discover such objects. The number of asteroids with known
satellites it is expected to be increased significantly due to the huge
amount of data expected from surveys like \textit{Gaia}. To that end, it
is necessary to develop automated strategies to find binary candidates
in such large datasets. 

\begin{figure}
\centering
\includegraphics[width=11cm]{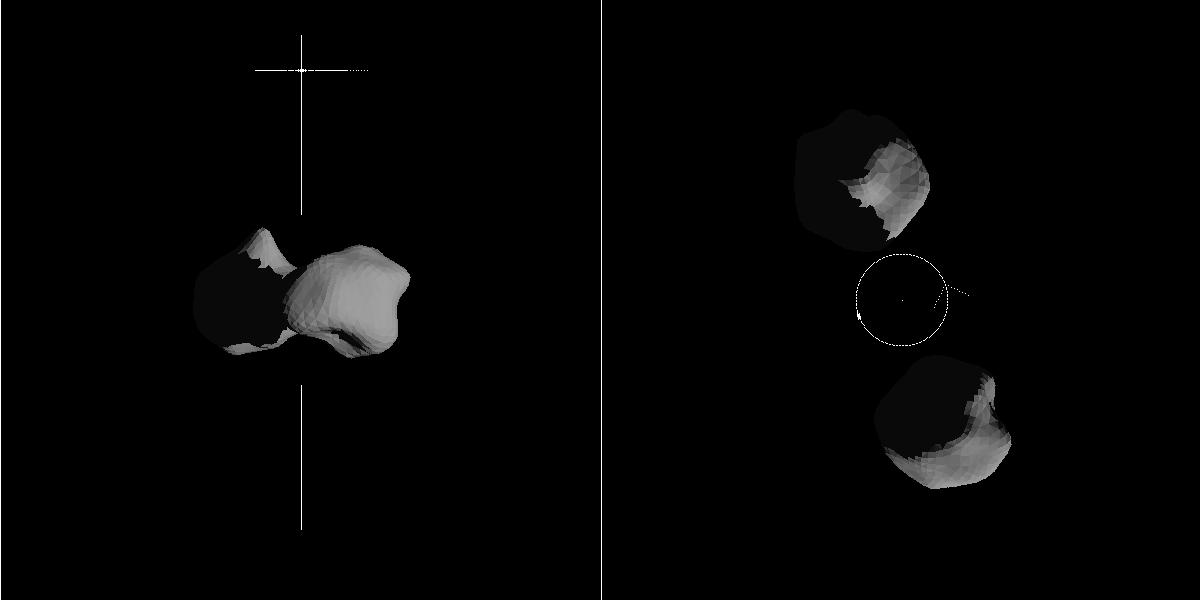}
\caption{Two different spatial views of the nonconvex model for 90 Antiope binary system shown at equatorial viewing (on the left), and the pole-on view on the right (model from Bartczak et al. \cite{BartczakAntiope}).}
\label{antiope}
\end{figure}

It is thought that NEO population can contain a high number of binary (or multiple) asteroids (see for instance Pravec \& Harris \cite{ph}).  One possible explanation for a high rate of multiplicity among this population could be the catastrophic disruption of rubble piles due to YORP spin up. Thus, inversion techniques capable of deriving models of binary asteroids can help us to better understand their formation processes and the physics of nongravitational effects.

\section{Models from various data types}
\label{sec:4}

Models derived from disk-integrated relative photometry can be combined
with other data types in order to derive additional physical properties
of asteroids. For instance, radar echo is a very powerful technique to
study the NEO population, while stellar occultation is an affordable
technique to obtain sizes of main belt (or even trans-neptunian!)
objects and discover satellites. In this chapter we briefly describe
both techniques and we outline a method to combine them with
lightcurves. Finally, it is worth noting that this is not intended to be
an extensive and comprehensive review of the data types that can
complement photometric observations. Other techniques like adaptive
optics, spectroscopy, thermal infrared observations or polarimetry
can be also combined with photometry to investigate asteroids. 
Here we provide the description of joining lightcurve data with radar imaging and with stellar occultation timings.

\subsection{Photometry and radar}
\label{sec:radar}

An Asteroid radar image is a reconstruction of a radio signal sent from the Earth and reflected by body's
surface. For this reason, this technique is best working for objects
approaching the Earth, such as NEOs. One dimension of such image comes
from time delay, as the signal has to travel different distances depending on which
a part of asteroid's surface it is reflected from. Second dimension
is directly associated with body's rotation. Received echo's
frequency is shifted with respect to incident ray due to Doppler effect
and depends on radial velocity of a surface element which increases as
we move away from asteroid's rotation axis. Range of frequency shift
depends on asteroid's rotation period and aspect (angle between rotation
axis and direction to the observer). As a result one can produce
range-Doppler image where each pixel value corresponds to echo power at
certain distance and radial velocity.

	Radar techniques can only be used when precise astrometry is available.
	Importance of knowing body's orbit accurately, especially in
	case of NEOs, cannot be overestimated. 
	
	Radar imaging is a rare situation in astronomy where we conduct
	an actual experiment by controlling the signal that gets
	reflected from a target body. It becomes possible to probe
	asteroids surface features comparable in size with signal
	wavelength or even have a glance at sub-surface properties. 
	Asteroid's shape is represented on images derived from
	radar observations in a form of a blend of top and bottom view of
	asteroid (in respect to line of sight). By examining radar
	images astronomers can determine large scale surface features
	such as big concavities or adjudge whether an object has
	satellites. Body size constrains can also be obtained. 
	
	Radar images are a very rich source of
	information and are used to create accurate three dimensional
	asteroid models that consist of asteroid's shape, spin axis
	orientation and rotation period. It can be done by using radar
	data only or by combining them with photometric data. Shape
	program \cite{betulia, improved_shape} is broadly, if not the only,
	such algorithm in use. It is an iterative method that starts with
	triaxial ellipsoid and by gradually changing initial shape and
	spin parameters arrives at a final model. Both radar and optical
	scattering laws are assumed prior to modelling process and stay
	fixed.

\begin{figure}
		\begin{center}
			\includegraphics[width=11cm]{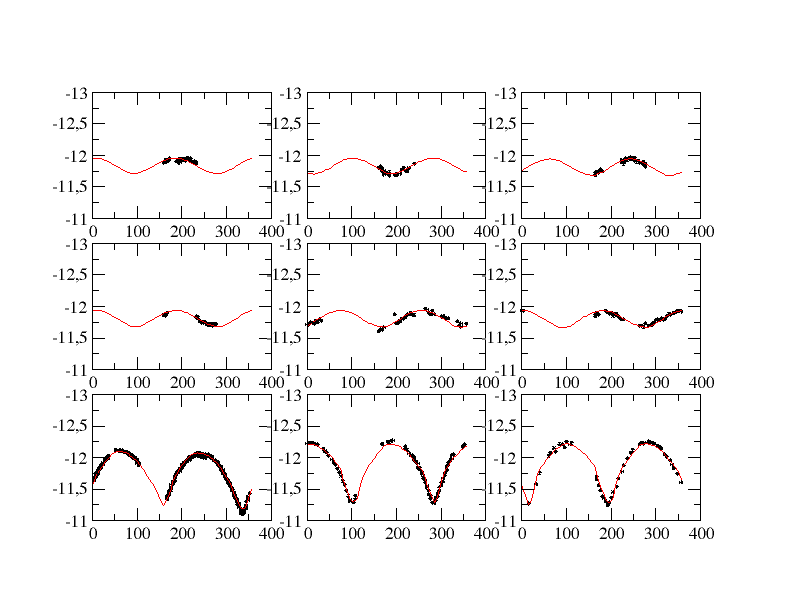}
		\end{center}
		\caption{ Selected light curves of 1996 HW1. Black points represent observational data while solid line represent synthetic lightcurves modelled with SAGE.  }
		\label{fig:becker_optical}
	\end{figure}

\begin{figure}
		\begin{center}
			\includegraphics[width=11cm]{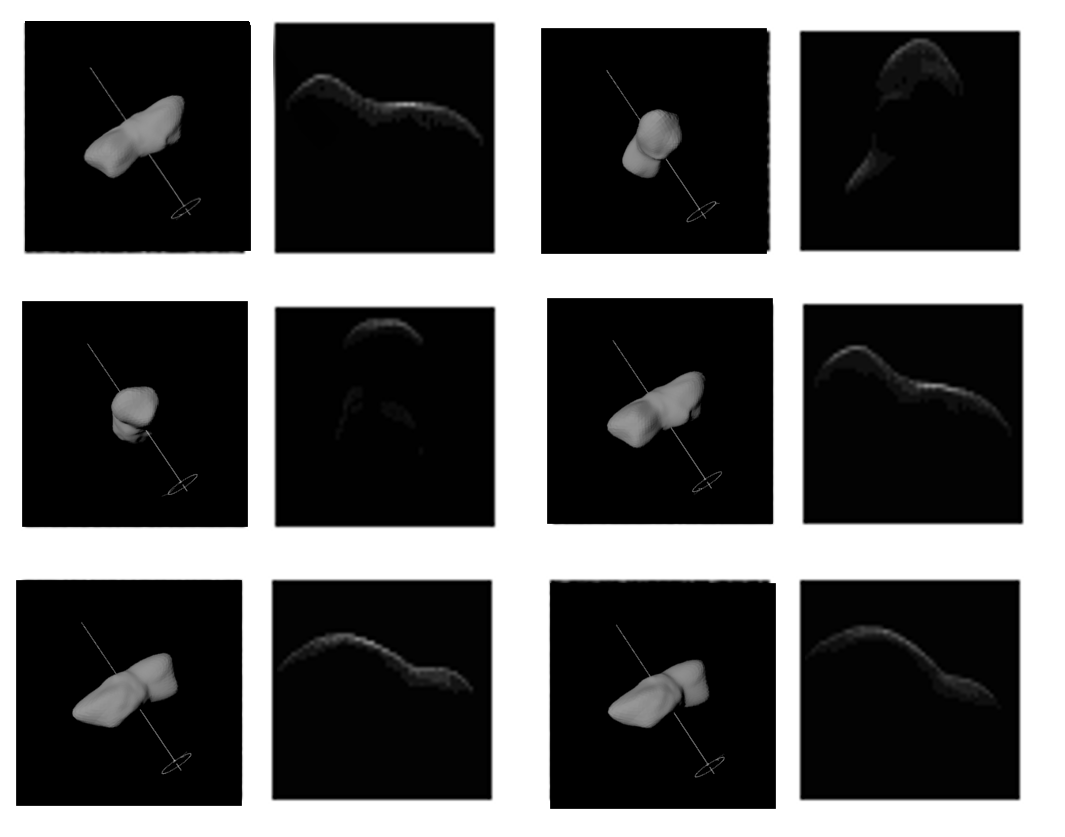}
		\end{center}
		\caption{ Comparison of range-Doppler images of 1996 HW1. First and third column:
simulated asteroid model as seen by observer; second and last column: simulated range-
Doppler image based on a SAGE model.  }
		\label{fig:becker_radar}
	\end{figure}

	In each iteration a simulated asteroid image is created and
	compared with the observations.  If radar data is used alongside
	photometric data, both images are computed separately from the
	same model. Every point on synthetic lightcurve is a calculated
	amount of light reflected by model's surface
	(Fig.~\ref{fig:becker_optical}, Fig.~\ref{fig:becker_radar}).
	Similarly, a radar range-Doppler image is created. Then a $\chi^2$ is determined by
	minimizing the differences between synthetic and observational
	measurements. Data is weighted depending on the type and quality, 
	and additional penalty functions defined by user are taken into
	consideration. This approach steers modelling process into
	global minimum.  Final model has to fit both radar and
	photometric observations. 

	The shape of an asteroid is derived in three stages and best fitting 
	model from previous stage becomes initial model of the next.
	The first stage uses only triaxial ellipsoid as a shape model; in the second
	stage, model is represented by spherical harmonics to be then
	transformed into polyhedral model described by vertices and
	triangular surface elements in the last stage. Concavities are
	allowed only at the last stage as they are difficult to
	represent using spherical harmonics (especially those of a low degree). 
	
	The described method is not fully automatic, meaning that it needs considerable
 human interference. Modelling is initialized with different
	values for parameters that are being further changed during the
	process to help the algorithm to come up with the model that
	reflects all important features present in observational data. 
	This has it's reflection in the low number of
	modelled asteroids based on radar observations.

\subsection{Photometry and stellar occultations}
\label{sec:occultation}	

\begin{figure}
		\begin{center}
			\includegraphics[width=11cm]{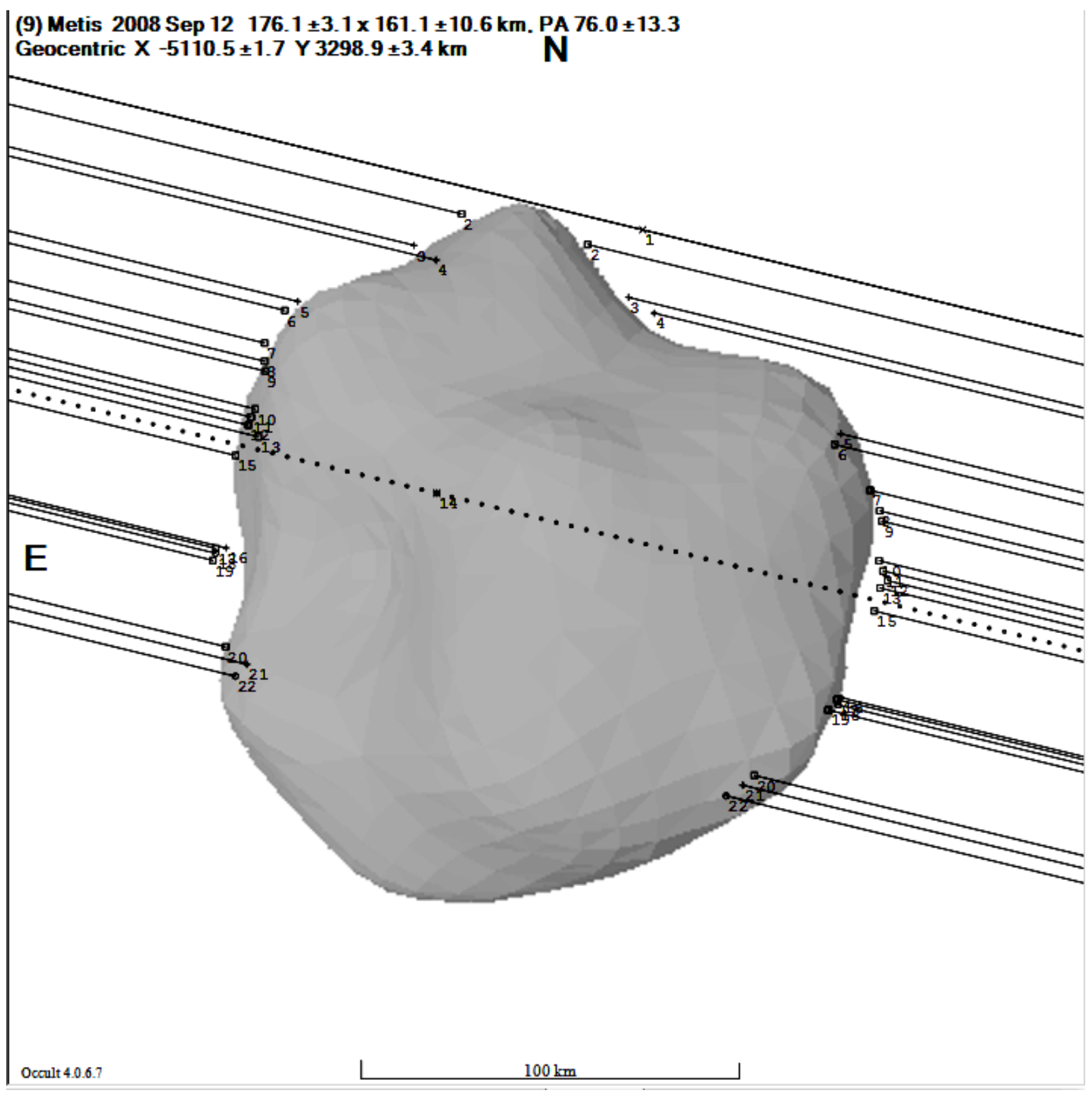}
		\end{center}
		\caption{ Best fit of a SAGE nonconvex model of (9) Metis \cite{metis} to the stellar occultation chords obtained during the 2008 occultation \cite{timerson}.}
		\label{fig:metis}
	\end{figure}

Stellar occultation is a interesting technique of asteroid imaging. The
idea is to measure asteroid's shadow cast on Earth's surface while it
passes in front of a star. Observers are set on shadow path and each of
them notes a time of the beginning and end of occultation (e.g.
Fig.~\ref{fig:metis}).

	In order to predict a shadow path and allow observers to choose
	observing sites properly, an asteroid's ephemeris as well as
	an occulted star position have to be very precise. Only if
	the ephemeris and star position are known with high accuracy
	will occultation time measurements result in good coverage of
	the body's shape. Nowadays, this represents a constraint as the
	availability of precise star astrometry is limited to the
	Hipparcos catalogue. Publication of \textit{Gaia} catalogue will
	greatly improve occultation events predictions; accuracy that is now 
	available for main belt asteroids 100 km in diameter, will be
	achievable for 15 km asteroids \cite{tanga_gaia}. On the other
	hand, this technique is not appropriate to NEOs, as their fast
	apparent movement against the background of distant stars
	constrict the possibility of recording the event. 
	
Despite this fact, this technique has been included in this paper, as it allows for investigations of internal structure of main belt objects. 
Keeping in mind that they are the source of asteroids with Earth-crossing orbits, these studies also increases our knowledge about NEOs composition.

	Choosing observational sites is crucial to successful
	occultation time measurements. Ideally, observers will cover the
	path of the shadow evenly and densely. Shadow path prediction
	however is directly dependent on the knowledge of ephemeris of
	the observed Solar System body and an estimation of the body's
	size. With more precise stellar catalogues, like the one ESA
	\textit{Gaia} mission will produce, more occultation events will be
	predicted with better precision, even for small bodies.
	
	Occultation observations are mainly carried out by amateur astronomers.
	Observing groups have to be mobile in order to cover the right
	area on the ground; fortunately stellar occultation events can
	be observed with small telescopes if occulted star is a bright
	one. Systems like GPS are of great help when it comes to
	establishing the observer's location and time essential to valid
	measurements. 
		
	Every chord (line along the shadow path) with marked beginning
	and end of the occultation event provides two points on the
	plane of sky. Given many points (five at minimum,
	\cite{occultations_elliot}) an ellipse can be fit to match these
	points. This is a good estimate of shape for large bodies, e.g.
	planets and TNOs objects; in case of smaller bodies this method
	provides rough estimate only as asteroids take a wide range of
	convex shapes that are irregular in general. 
	
	Stellar occultation remains the best available method for
	determining body's radius its resolution being at the level of
	kilometres. Given dense chord distribution the resulting body's
	shape envelope is very precise thus fit for verifying models
	obtained by inversion techniques against reality.
	
	Models can be enriched with additional information, the size of
	the body being the most valuable one as it gives the model
	proper dimensions and allows density and albedo estimations, as was done in case of (90) Antiope
	(see for instance Bartczak et al. \cite{BartczakAntiope}).
	The described method is also sufficient to determine whether an
	asteroid is a binary system.

	Analysis of lightcurve profiles captured during occultation
	event (immersion and emersion at the beginning and end of
	occultation) can tell us about the presence of an atmosphere
	\cite{occultations_elliot}. Moreover, rings around bodies can
	by detected and studied, like in the case for
	giant planets in Solar System. It is possible to detect rings
	around smaller objects \cite{chariklo} but it requires extremely
	high precision timings. Still it shows the power of the method, where some features
	could not be discovered or that precisely measured using other observing techniques.

\section{Conclusions}

    We have shown that asteroid modelling is an effort that needs to be undertaken in order to study the nongravitational forces affecting these small bodies. 
Disk-integrated photometry is the main technique used to derive spin states and shape models. However, for solving the inversion problem, lightcurves need to 
fulfil certain requirements in terms of quality and viewing geometries.

    A substantial part of this paper has been devoted to describe how to gather such data and what are the inversion techniques commonly applied. 
The first and simplest solution -- which is, however, still useful in some cases due to specific situations, such as the the analysis of the huge amount of data 
generated by the \textit{Gaia} mission -- is a triaxial ellipsoid representation of the body shape, for which synthetic brightness can be evaluated analytically, 
making it ideal for problems with high CPU demand. The so-called lightcurve inversion method, which solution consist of a convex shape model of the asteroid and 
its spin state, is a worthwhile technique when we are specifically interested in the study of the spin rate and shape outline of the body. 
However, other techniques producing more detailed shape solutions (i.e. nonconvex shapes) are necessary when investigating further physical properties. In particular, 
modelling of nongravitational effects acting on asteroids is extremely sensitive to the used shape representation. Some inversion methods are also able to derive 
shape models of binary asteroids. We have shown the example of SAGE, which is a technique capable of deriving nonconvex models for synchronous binary asteroids from relative photometry only.

    It is also possible to combine other observing techniques to derive additional physical properties of asteroids. In chapter four, we have described two techniques -- 
radar imaging and stellar occultation timings -- which are mainly used to scale the model in size and, in some cases, derive additional information. This includes the study of 
fine details in the shape or the discovery of satellites. In addition, we have briefly described some procedures to combine these observations with photometry in the modeling process.

\end{document}